\documentstyle[12pt,epsfig]{article}
\newcommand\mysection{\setcounter{equation}{0}\section}

\def\MSbar{\relax\ifmmode\overline{\rm MS}\else{$\overline{\rm MS}${ }}\fi}

\def\fun#1#2{\lower3.6pt\vbox{\baselineskip0pt\lineskip.9pt
  \ialign{$\mathsurround=0pt#1\hfil##\hfil$\crcr#2\crcr\sim\crcr}}}

\def\eV{{\rm e\kern-0.12em V}}            
  
\def\half{{\textstyle {1\over2}}}
  \def\quart{{\textstyle {1\over4}}}

\def \al {\relax\ifmmode{\alpha}\else{$\alpha${ }}\fi}
\def \be {\relax\ifmmode{\beta}\else{$\beta${ }}\fi}

\def\sub#1#2{{#1}_{\mbox{\scriptsize{#2}}}}
\def\beq{\begin{equation}}   \def\eeq{\end{equation}}
\def \as{\relax\ifmmode\alpha_s\else{$\alpha_s${ }}\fi}
\def \pt{\relax\ifmmode{p_t}\else{$p_t${ }}\fi}

\def\eps{\relax\ifmmode\epsilon\else{$\epsilon${ }}\fi}
\def\ee{\relax\ifmmode{e^+e^-}\else{${e^+e^-}$}\fi}
\def\qq{\relax\ifmmode{q\overline{q}}\else{$q\overline{q}${ }}\fi}
\def\pert{{\mbox{\scriptsize pert}}}
\newskip\humongous \humongous=0pt plus 1000pt minus 1000pt
\def\caja{\mathsurround=0pt}
\def\eqalign#1{\,\vcenter{\openup1\jot
\caja   \ialign{\strut \hfil$\displaystyle{##}$&$
\displaystyle{{}##}$\hfil\crcr#1\crcr}}\,}

\newif\ifdtup

\def\eqal2#1{\,\vcenter{\openup1\jot
\caja   \ialign{\strut \hfil$\displaystyle{##}$&\hfil$
\displaystyle{{}##}$\hfil &$
\displaystyle{{}##}$\hfil\crcr#1\crcr}}\,}

\def\etal{{\em et al.}}

\def\cpc#1#2#3{{\em Comp.\ Phys.\ Commun.\ }~\underline{#1} (19#3) #2}
\def\ar#1#2#3{{\em Ann.\ Rev.\ Nucl.\ Part.\ Sci.\ }~\underline{#1} (19#3) #2}
\def\ib#1#2#3{{\em ibid.\ }~\underline{#1} (19#3) #2}
\def\np#1#2#3{{\em Nucl.\ Phys.\ }~\underline{B#1} (19#3) #2}
\def\pl#1#2#3{{\em Phys.\ Lett.\ }~\underline{#1B} (19#3) #2}
\def\pr#1#2#3{{\em Phys.\ Rev.\ }~\underline{D#1} (19#3) #2}

\def\prl#1#2#3{{\em Phys.\ Rev.\ Lett.\ }~\underline{#1} (19#3) #2}

\def\sjnp#1#2#3{{\em Sov.\ J.\ Nucl.\ Phys.\ }~\underline{#1} (19#3) #2}
\def\spj#1#2#3{{\em Sov.\ Phys.\ JETP}\/~\underline{#1} (19#3) #2}

\def\zp#1#2#3{{\em Zeit.\ Phys.\ }~\underline{C#1} (19#3) #2}

 \def\cite#1{[\ref{#1}]}
 \def\citd#1#2{[\ref{#1},\ref{#2}]}
 \def\citt#1#2#3{[\ref{#1},\ref{#2},\ref{#3}]}
 \def\citq#1#2#3#4{[\ref{#1},\ref{#2},\ref{#3},\ref{#4}]}
 \def\citm#1#2{[\ref{#1}--\ref{#2}]}
\def\Apr{A^\prime}

\def\cF{{\cal{F}}}
\def\cO{{\cal{O}}}
\def\cR{{\cal{R}}}
\def\at{\al_{\mbox{\scriptsize eff}}}

\def\Li{\mbox{Li}_2}
\def\cFv{{\cal{F}}^{(v)}}
\def\re#1{(\ref{#1})}

\begin{document}
\begin{titlepage}
\begin{flushright}
Cavendish-HEP-96/9\\
hep-ph/9608394
\end{flushright}              
\vspace*{\fill}
\begin{center}
{\Large \bf\boldmath
Power Corrections and Renormalons in\\[1ex]
\ee\ Fragmentation Functions\footnote{Research supported in
part by the U.K. Particle Physics and Astronomy Research Council and by
the EC Programme ``Training and Mobility of Researchers", Network
``Hadronic Physics with High Energy Electromagnetic Probes", contract
ERB FMRX-CT96-0008.}}
\end{center}
\par \vskip 5mm
\begin{center}
        M.\ Dasgupta and B.R.\ Webber \\
        Cavendish Laboratory, University of Cambridge,\\
        Madingley Road, Cambridge CB3 0HE, U.K.
\end{center}
\par \vskip 2mm
\begin{center} {\large \bf Abstract} \end{center}
\begin{quote}
We estimate the power corrections (infrared renormalon contributions)
to the coefficient functions for the transverse, longitudinal and
asymmetric fragmentation functions in \ee\ annihilation, using
a method based on the analysis of one-loop Feynman graphs
containing a massive gluon.
The leading corrections have the expected $1/Q^2$ behaviour, but
the gluonic coefficients of the longitudinal and transverse
contributions separately have strong singularities at small $x$,
which cancel in their sum. This leads to $1/Q$ corrections to
the longitudinal and transverse parts of the annihilation
cross section, which cancel in the total cross section.
\end{quote}
\vspace*{\fill}
\begin{flushleft}
     Cavendish--HEP--96/9\\
     August 1996
\end{flushleft}
\end{titlepage}

\mysection{Introduction}

Experimental studies of the single-hadron inclusive spectrum
in the \ee\ annihilation process $\ee\to\gamma^*/Z^0\to hX$
have been performed with high precision over a wide range of
energies for various types of produced hadrons $h$
(see Refs.~\citm{jp_scaTASSO}{jp_longOPAL}
and references therein). These studies have mostly concerned
the total fragmentation function,
\beq\label{sigtot}
\frac{1}{\sub\sigma{tot}}\frac{d\sigma}{dx}(\ee\to hX)
\equiv\sub F{tot}(x,Q^2)\;,
\eeq
where $x=2Q\cdot p_h/Q^2$ is the energy fraction of the
observed hadron and $Q^\mu$ is the virtual boson
four-momentum, equal to the overall centre-of-mass momentum
in this process. The most precise data are those for the
total fragmentation function into charged hadrons.
The scaling violations (logarithmic $Q^2$-dependence)
in this quantity predicted by perturbative QCD
\citm{jp_DGLAP}{jp_NWa} have been observed and used
to measure the strong coupling constant
\citd{jp_scaDELPHI}{jp_scaALEPH}.

The ALEPH \cite{jp_scaALEPH} and OPAL \cite{jp_longOPAL}
collaborations at LEP have also presented results on the
joint distribution in the energy fraction $x$ and the
angle $\theta$ between the observed hadron $h$ and the
incoming electron beam. We can write \cite{jp_NWa}
\beq\label{sigTLA}
\frac{1}{\sub\sigma{tot}}\frac{d^2\sigma}{dx\,d\cos\theta}
= \frac 3 8 (1+\cos^2\theta)\,\sub F T(x,Q^2)
+ \frac 3 4    \sin^2\theta \,\sub F L(x,Q^2)
+ \frac 3 4    \cos  \theta \,\sub F A(x,Q^2)\; ,
\eeq
where $\sub F T$, $\sub F L$ and $\sub F A$ are respectively the
transverse, longitudinal and asymmetric fragmentation functions.
As their names imply, $\sub F T$ and $\sub F L$ represent the
contributions from virtual bosons polarized transversely or
longitudinally with respect to the direction of motion of the
observed hadron.  $\sub F A$ is a parity-violating contribution
which comes from the interference between the $Z^0$ and photon
contributions.  Integrating over all angles, we obtain the
total fragmentation function \re{sigtot},
$\sub F{tot}=\sub F T +\sub F L$.

According to the factorization theorems of QCD \cite{jp_CS},
each of the functions $\sub F P$ (P = tot,T,L or A) can be represented
as a convolution of universal parton fragmentation functions $D_i$
($i=q,\bar q$ or $g$) with perturbatively calculable coefficient functions
$\sub C P^i$ \citm{jp_CFP}{jp_NWa}:
\beq\label{convol}
\sub F P(x,Q^2) = \sum_i\int_x^1\frac{dz}{z} \sub C P^i(z)\,D_i(x/z,Q^2)\;.
\eeq
The parton fragmentation functions themselves cannot be computed
perturbatively, although their logarithmic $Q^2$-dependence, which is
the main source of scaling violation, is predicted by the
QCD evolution equations \cite{jp_DGLAP}. The evolution kernels
(splitting functions) differ in non-leading orders \citd{jp_CFP}{jp_FP}
from those for deep inelastic structure functions.  Thus scaling
violation in fragmentation provides an important independent test
of QCD.  The longitudinal and transverse fragmentation functions can
be used to measure the gluon fragmentation function
\citd{jp_scaALEPH}{jp_longOPAL}, while the asymmetric part may be useful
for the measurement of electroweak couplings \cite{jp_NWb}. Once
measured and parametrized \citq{jp_scaALEPH}{jp_NWa}{GreRol}{BinKK},
the parton fragmentation functions can also be used to test QCD
in jet fragmentation in other processes such as lepton-hadron
\cite{jp_lp} and hadron-hadron \cite{jp_had} collisions.

In addition to the logarithmic $Q^2$-dependence predicted by
perturbative QCD, it is expected that fragmentation functions
will exhibit process-dependent
power corrections, i.e.\ contributions proportional
to inverse powers of $Q$, analogous to the higher-twist
contributions found in deep inelastic scattering \cite{jp_ht}.
An understanding of power corrections is crucial for
precision tests of scaling violation.
In the deep inelastic case such contributions are related
to the hadronic matrix elements of local operators, and it
is well established that the leading corrections should be
of order $1/Q^2$.  The machinery of the local operator
product expansion is not applicable to fragmentation, but
a study of the relevant non-local operator matrix elements
again suggests a $1/Q^2$ behaviour \cite{jp_BalBra}.
The current data on the total \ee\ fragmentation
function $\sub F{tot}$ are consistent with either
a $1/Q$ or $1/Q^2$ form for the leading power correction
\cite{jp_scaALEPH}.
Data on the separate functions $\sub F P$ for P=T, L
and A are at present limited to a single energy
$Q=M_Z$ \citd{jp_scaALEPH}{jp_longOPAL}, and so
predictions concerning their scaling violation and
power corrections remain to be tested.

In the present paper we estimate the power corrections to
fragmentation functions using a recently developed `dispersive'
method based on the infrared properties of Feynman graphs and some
assumptions about the strong coupling at low scales \cite{BPY}.
The techniques involved are similar to those involved in
the study of infrared renormalons \citm{renormalons}{lomax},
which correspond to unsummable divergences of the perturbative
expansion. Here we use them as a more general probe of the
influence of soft regions of integration on
hard process observables. The results obtained by
applying these techniques to deep inelastic scattering
\citt{BPY}{DasWeb96}{Stein} are consistent with those
of the operator product expansion and look promising
phenomenologically.

A calculation of the leading power correction to the total
fragmentation function $\sub F{tot}$ using the dispersive
approach was presented in Ref.~\cite{BPY}.
A $1/Q^2$ correction was found and the corresponding quark coefficient
function was computed. In the present paper we extend these results to
subleading ($1/Q^4$) power corrections, and to the transverse,
longitudinal and asymmetric quark fragmentation functions separately.
We also compute (in a certain approximation) the corresponding
gluonic coefficient functions.

We shall be particularly concerned to clarify an apparent paradox
which arises when one considers sum rules for fragmentation.
Summed over all particle types, the total fragmentation function
satisfies the energy sum rule, which we may write as
\beq
\frac 1 2\int_0^1 dx\,x\sub F{tot}(x,Q^2) = 1\;.
\eeq
Similarly the integrals
\beq
\frac 1 2\int_0^1 dx\,x\sub F {T,L}(x,Q^2) \equiv
\frac{\sub\sigma{T,L}}{\sub\sigma{tot}}
\eeq
give the transverse and longitudinal fractions of the total
cross section.  The perturbative prediction is \cite{jp_RvN} 
\beq\label{sigl}\eqalign{
\frac{\sub\sigma{L}}{\sub\sigma{tot}} =
1-\frac{\sub\sigma{T}}{\sub\sigma{tot}} &= \frac{\as}{\pi} +
\left(\frac{601}{40}-\frac{6}{5}\zeta(3)-\frac{37}{36}n_f\right)
\left(\frac{\as}{\pi}\right)^2+ O(\as^3) \cr
&\simeq \frac{\as}{\pi} +
(13.583-1.028 n_f)\left(\frac{\as}{\pi}\right)^2}
\eeq
where $\zeta(3)=1.202$ and $n_f$ is the number of active
flavours.  Note that the whole of the $\cO(\as)$
correction to $\sub\sigma{tot}$ comes from the longitudinal part,
while the $\cO(\as^2)$ correction receives both longitudinal and
transverse contributions.

\begin{figure}\begin{center}
\epsfig{figure=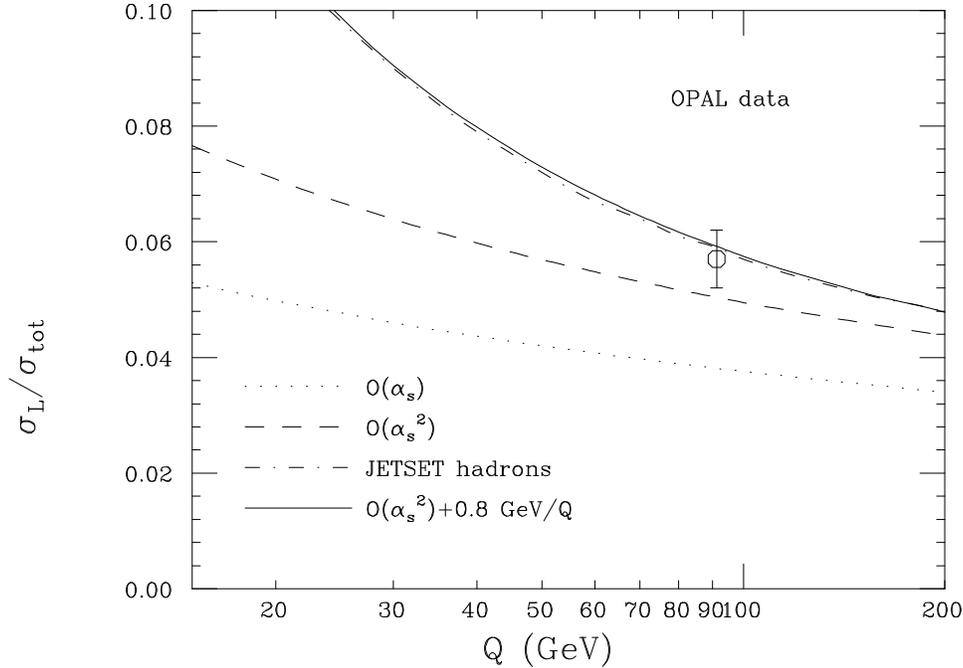,height=9.0cm}
\caption{Longitudinal fraction of the $\ee$ hadronic cross section.}
\label{opal_sigl}
\end{center}\end{figure}

The OPAL data point \cite{jp_longOPAL}
for $\sub{\sigma}{L}/\sub\sigma{tot}$ is shown,
together with the perturbative predictions, in Fig.~{\ref{opal_sigl}}.
The data lie somewhat above the next-to-leading-order prediction (dashed),
which suggests that higher-order and/or non-perturbative corrections
are significant. An estimate of the latter is provided by the difference
between the JETSET Monte Carlo \cite{JETSET} hadron-level prediction
(dot-dashed) and the perturbative result. This difference shows a clear
$1/Q$ behaviour, with a coefficient of about 1 GeV.  Assuming that the
same behaviour will be manifest in the data, we have an apparent
paradox: the function $x\sub F L$ has only a $1/Q^2$ correction, but
its integral has a $1/Q$ correction.

Power corrections proportional to $1/Q$,
usually referred to as hadronization corrections,
are typical of hadronic event shapes in \ee\ annihilation,
where they are clearly seen in the data and are predicted
by a variety of approaches \citm{tube}{jp_NaSe}.
String-like models of hadronization, for example,
in which the energy of a jet is redistributed with
proper density $\lambda\sim 500$ MeV per unit rapidity and
limited transverse momentum relative to the jet axis, imply
a correction to the longitudinal cross section of
\cite{jp_NWa}\footnote{Note that there is a misprint in
the corresponding equation (3.35) of Ref.~\cite{jp_NWa}.}
\beq\label{lhad}
\frac{\delta\sub{\sigma}{L}}{\sub\sigma{tot}}
= \frac{\pi\lambda}{2Q}\simeq \frac{0.8\;\mbox{GeV}}{Q}\;.
\eeq
Adding this to the next-to-leading-order prediction
gives the solid curve in Fig.~{\ref{opal_sigl}},  
which agrees well with the JETSET prediction.
The correction arises from mixing between the transverse and
longitudinal angular dependences in Eq.~\re{sigTLA} due
to hadronization.  The transverse cross section receives
an equal and opposite correction, and so there
is no $1/Q$ term in the total cross section.

We shall see that dispersive approach of Ref.~\cite{BPY}
leads to the following resolution of the paradox.
The $1/Q$ terms arise from soft gluon fragmentation.
The gluonic coefficient functions of the $1/Q^{2p}$ power
corrections are highly singular at small $x$. Upon
integration they are all `promoted' to a $1/Q$
behaviour and have to be resummed. The result is a $1/Q$ correction
to $\sub\sigma{L}/\sub\sigma{tot}$, with a coefficient similar to
that in the hadronization models.

In the remainder of the paper, we first give a brief summary of the
relevant assumptions and results from Ref.~\cite{BPY}. The
dispersive method is based on the evaluation of one-loop
Feynman graphs containing a gluon of finite mass $\mu$.
This yields the `characteristic functions' for the relevant
quantities, which are given in Sect.~3. We extract the power
corrections from the behaviour of the characteristic functions
as $\mu\to 0$. The rules for taking this limit are also
explained in Sect.~3. In Sect.~4 we give the expressions
thus obtained for the power corrections to the fragmentation
functions, and to the transverse and longitudinal cross
sections. Finally in Sect.~5 we discuss the results and
give some numerical predictions.

\mysection{Dispersive method}

We do not repeat the discussion of Ref.~\cite{BPY} but
simply summarize the results required here.  The
basic assumption is that the dominant non-perturbative
contributions to the coefficient functions $\sub C P^i$
in Eq.~\re{convol} are of the form
\beq\label{deltaC}
\delta\sub C P^i(x,Q^2) = \int_0^\infty \frac{d\mu^2}{\mu^2}\, 
\delta\at(\mu^2)\dot\sub\cF P^i(x,Q^2;\mu^2)\,. 
\eeq
Here $\delta\at(\mu^2)$ is a non-perturbative modification to the
effective strong coupling, restricted to the region of low $\mu^2$.
$\sub\cF P^i(x,Q^2;\mu^2)$ is the relevant {\em characteristic function},
that is, the parton fragmentation function computed at order $\as$
using a non-zero value $\mu$ for the gluon mass in the Feynman
denominators of the contributing graphs. $\dot\sub\cF P^i$ is minus
the logarithmic derivative of $\sub\cF P^i$ with respect to $\mu^2$. 
Since $\sub\cF P^i$ depends only on dimensionless ratios, we may write
\beq
\sub\cF P^i(x,Q^2;\mu^2) =\sub\cF P^i(x,\eps)\;,
\;\;\;\;\;\;
\dot\sub\cF P^i(x,\eps) \equiv -\eps\frac{\partial}{\partial\eps}
\sub\cF P^i(x,\eps)\;,
\;\;\;\;\;\;
\eps\equiv\frac{\mu^2}{Q^2}\;.
\eeq
The non-perturbative contributions $\delta\sub C P^i$ thus
depend on the small-$\eps$ behaviour of $\sub\cF P^i$, which
is of the generic form
\beq\label{Fsmalleps}
\sub\cF P^i(x,\eps) = -\sub P P^i(x)\ln\eps + \sub C{0,P}^i(x)
- \sub C{2,P}^i(x)\eps\ln\eps
- \half \sub C{4,P}^i(x)\eps^2\ln\eps -\cdots\,,
\eeq
where the dots indicate terms that are either $\cO(\eps^3\ln\eps)$
or analytic and vanishing at $\eps=0$.  Here $\sub P P^i(x)$
(contributing for P=T,A only) is the $q\to i$ splitting function,
$\sub C{0,P}^i(x)$ is the relevant perturbative coefficient function
(in the gluon mass regularization scheme), and $\sub C{2,P}^i(x)$
etc.\ will be related to non-perturbative corrections.

A crucial point is that, for consistency with the
operator product expansion, the  integer $\mu^2$-moments
of the coupling modification should vanish:
\beq\label{vanish}
 \int_0^\infty \frac{d\mu^2}{\mu^2} \mu^{2p} \delta\at(\mu^2)\>=\>0\>, 
\eeq
at least for the first few moments $p=1,\ldots,p_{\max}\sim 9$.
As a consequence, only those terms in the small-$\eps$
behaviour of $\dot\sub\cF P^i(x,\eps)$
that are non-analytic at $\eps =0$ lead to
non-perturbative contributions \cite{BBB}.
Thus from the small-$\eps$ behaviour \re{Fsmalleps} we find
\beq\label{CNP}
\delta\sub C P^i(x,Q^2) = \sub C{2,P}^i(x)\frac{\Apr_2}{Q^2}
+ \sub C{4,P}^i(x)\frac{\Apr_4}{Q^4}+\cdots\;,
\eeq
where, following Ref.~\cite{BPY}, we have defined the log-moment
integrals
\beq\label{adefs}
\Apr_{2p} \>=\>\frac{C_F}{2\pi}
\int_0^\infty\frac{d\mu^2}{\mu^2}
\>\mu^{2p}\,\ln(\mu^2/\mu_0^2)\>\,\delta\at(\mu^2)\;.
\eeq
Notice that since integer $\mu^2$-moments of $\delta\at$ vanish,
these quantities are independent of the scale $\mu_0^2$.  For
convenience, we extract a universal factor of $C_F/2\pi$
from the characteristic function.

Instead of interpreting the coefficients $\Apr_{2p}$ in
terms of a universal low-energy effective coupling, one
may treat them more generally as process-dependent
parameters to be determined experimentally. Eq.~\re{CNP} still
has predictive power because the coefficient functions
$\sub C{2,P}^i(x)$ etc.\ specify the $x$-dependence of the
power corrections. This dependence is supposed to reflect the
relative sensitivity of different regions of $x$ to soft dynamics.

The technique we use to evaluate the coefficient functions of
power corrections, viz.\ extraction of the non-analytic terms
in the massive-gluon expressions for observables as $\mu^2\to 0$,
is the same as that applied in studies of infrared
renormalons \cite{BBB}. In the language of renormalons,
the terms computed are ambiguities in the perturbative
prediction for the observable in question, which have
to cancel against corresponding ambiguities in
power-suppressed non-perturbative contributions.
Here we argue that the non-perturbative contributions
themselves should display the same power behaviour
and $x$-dependence, since the small-$\mu^2$ limit
probes the sensitivity of an observable to
the soft non-perturbative region as a function
of $x$ and $Q^2$.

We expect quark and gluon fragmentation to contribute to
power corrections on an equal basis, since the dominant
contributions are assumed to be determined at first
order in $\at$.  The application of the dispersive method
to compute the quark contribution is straightforward
since in that case we sum inclusively over
all gluon fragmentation products. The inclusive sum
generates a contribution which is equivalent to that
of a massive gluon, as discussed in ref.~\cite{BPY}.
We shall also use the dispersive approach to calculate gluonic
contributions, but its application there is more questionable.
By definition we observe the fragmentation products
of the gluon in that case, which spoils the
equivalence to a massive gluon. The situation
becomes similar to that for an event shape variable:
the correction has the same leading power behaviour
as that due to a massive gluon, but the coefficient
may be modified. This can be investigated in the
large-$n_f$ limit, in which the gluon fragments
only into quark-antiquark pairs \citd{jp_NaSe}{BBprep}.
Although the question requires further study, we
assume here that the massive-gluon technique does
provide a reasonable estimate of contributions from
gluon as well as quark fragmentation.

\mysection{Characteristic functions}

The object of central importance in the dispersive method is the
characteristic function $\sub\cF P^i(x,\eps)$ for the emission of a
gluon with mass-squared $\mu^2 = \eps Q^2$ at the hard scale $Q^2$.
This is computed from the relevant one-loop graphs with a modified
gluon propagator.  The characteristic function for the
total fragmentation function is obtained by
contracting the resulting hadronic tensor with the tensor
representing a sum over virtual-boson polarization states,
\beq
\sum_{\mbox{\scriptsize P}}\sub\varepsilon P^\mu\sub\varepsilon P^\nu
= -g^{\mu\nu} + \frac{Q^\mu Q^\nu}{Q^2}\;.
\eeq
The corresponding tensor for the longitudinal part is
$\sub\varepsilon L^\mu\sub\varepsilon L^\nu$ where
$\sub\varepsilon L^\mu$ is the polarization vector
along the direction of $\mbox{\bf p}_h$, the
three-momentum of the observed hadron
in the virtual-boson rest frame:
\beq
\sub\varepsilon L^\mu = \frac{1}{|\mbox{\bf p}_h|}\left(p_h^\mu
- \frac{Q\cdot p_h\, Q^\mu}{Q^2}\right)\;.
\eeq
The characteristic function for the transverse fragmentation
function is then obtained as the difference between the total
and the longitudinal part. For the
asymmetric fragmentation function we take the vector-axial
interference term and contract with the tensor
\beq
-\frac i 2\eps_{\mu\nu\sigma\rho}\;
\frac{p_h^\sigma Q^\rho}{|\mbox{\bf p}_h|\,Q}\;.
\eeq
The resulting expressions for $\sub\cF P^i(x,\eps)$ are given below.
The power corrections are deduced from the non-analytic
terms in the small-$\eps$ behaviour of the
logarithmic derivative $\dot\sub\cF P^i(x,\eps)$. When taking the
small-$\eps$ limit, one should be careful with the phase-space boundaries,
and in particular with functions that are singular on or near these
boundaries. We give the rules for obtaining the correct limiting
behaviour.

\subsection{Quark fragmentation}

The phase space for fragmentation into a quark of negligible mass
with emission of a gluon of mass-squared $\eps Q^2$ is $0<x<1-\eps$.
The characteristic function is therefore of the form
\beq\label{FTq}
\sub\cF P^q(x,\eps) = \sub\cF P^{(r)}(x,\eps)\,\Theta(1-x-\eps)
+\cFv(\eps)\,\delta(1-x)
\eeq
where $\cFv$ represents the virtual contribution.
For transverse quark fragmentation the real gluon emission part is
\beq\label{FTr}\eqalign{
\sub\cF T^{(r)}(x,\eps) &= \left[\frac {2(1+\eps)^2}{1-x}-1-x
-2\eps +4\frac{\eps}{x}+6\frac{\eps^2}{x^2}\right]
\ln\left[\frac{(x+\eps)(1-x)}{\eps}\right] 
\cr&
-\frac{3-5\eps^2}{2(1-x)} +\frac{\eps}{(1-x)^2}
+\frac{\eps^2}{2(1-x)^3}+\frac{\eps}{x+\eps}
-6\frac{\eps(1-\eps)}{x}-\frac{1}{2}(1-x)+3\eps\;.
}\eeq
The virtual contribution is
\beq\label{Fv}
\cFv(\eps) =
2(1+\eps)^2\left[\Li(-\eps)+\ln\eps\ln(1+\eps)
-\frac 12\ln^2\eps +\frac{\pi^2}{6} \right]  
-\frac72 -(3+2\eps)\ln\eps-2\eps 
\,,
\eeq
where 
\beq
\Li(-\eps)= -\int_0^\eps\frac{dt}{t}\ln(1+t)\,.
\eeq

The characteristic function for longitudinal quark fragmentation
has only a contribution from real gluon emission, which is
\beq\label{FLr}\eqalign{
\sub\cF L^q(x,\eps) &= -2\frac{\eps}{x}\left(2+3\frac {\eps}{x}\right)
\ln\left[\frac{(x+\eps)(1-x)}{\eps}\right] 
\cr&
-2\frac{\eps(1+2\eps)}{1-x} +\frac{\eps^2}{(1-x)^2}
+6\frac{\eps(1-\eps)}{x}+1-2\eps\;.
}\eeq

The characteristic function for asymmetric quark
fragmentation is of the form \re{FTq}
with the same virtual contribution, but the
real gluon emission part becomes
\beq\label{FAr}\eqalign{
\sub\cF A^{(r)}(x,\eps) &= \left[\frac {2(1+\eps)^2}{1-x}-1-x
-2\eps +2\frac{\eps(2+\eps)}{x}\right]
\ln\left[\frac{(x+\eps)(1-x)}{\eps}\right] 
\cr&
-\frac{3(1-\eps^2)}{2(1-x)} +\frac{\eps}{(1-x)^2}
+\frac{\eps^2}{2(1-x)^3}-\frac{\eps}{x+\eps}
-\frac{3}{2}(1-x)+\eps\;.
}\eeq

In taking the small-$\eps$ limits of Eqs.~\re{FTr}, \re{FLr} and
\re{FAr}, we must remember that the phase-space boundary
is at $x=1-\eps$.  Now for any function $F(x)$ that is analytic in a
neighbourhood of $x=1$ and any test function $f(x)$, we have
\beq
\int_0^{1-\eps} F(x)\,f(x)\,dx
=\int_0^1 F(x)\,f(x)\,dx - \eps F(1)f(1) +\half\eps^2
[F'(1)f(1)+F(1)f'(1)] + \cdots\;.
\eeq
Recalling that
\beq\label{deln}
\int_0^1 \delta^{(n)}(1-x)\,f(x)\,dx = f^{(n)}(1)\;,
\eeq
we can make the $\eps$-dependence explicit, up to terms of
order $\eps^2$, by replacing any expression
$F(x)$ analytic at $x=1$ as follows:
\beq\label{regsmalleps}
F(x)\to F(x)-\eps [F(1)-\half\eps F'(1)]\,\delta(1-x)
+\half\eps^2 F(1)\,\delta'(1-x)\;.
\eeq

For expressions that are singular at $x=1$, we define `+',
`++' and `+++' prescriptions such that, for any test function $f(x)$,
\beq\eqalign{
  \int_0^1 F(x)_+\,f(x)\,dx &= \int_0^1 F(x)\,[f(x)-f(1)]\,dx\,\cr
  \int_0^1 F(x)_{++}\,f(x)\,dx &= \int_0^1
F(x)\,[f(x)-f(1)+(1-x)f'(1)]\,dx\,\cr
  \int_0^1 F(x)_{+++}\,f(x)\,dx &= \int_0^1
F(x)\,[f(x)-f(1)+(1-x)f'(1)-\half(1-x)^2 f''(1)]\,dx\;.
}\eeq
Using Eq.~\re{deln}, we can now
replace the singular terms, up to terms of order $\eps^2$,
as follows:\footnote{Note that these rules are different from
those given in Ref.~\cite{DasWeb96} for the deep inelastic case,
since the phase space is different.}
\beq\label{smalleps}\eqalign{
  \frac{1}{1-x} &\to \frac{1}{(1-x)_+}
-\ln\eps\,\delta(1-x) +\eps\,\delta'(1-x)-\quart\eps^2\,\delta''(1-x) \cr
  \frac{\ln(1-x)}{1-x} &\to \left(\frac{\ln(1-x)}{1-x}\right)_+
-\half\ln^2\eps\,\delta(1-x)+\eps(\ln\eps-1)\,\delta'(1-x)
 -\quart\eps^2(\ln\eps-\half)\,\delta''(1-x)\cr
  \frac{\eps}{(1-x)^2} &\to \frac{\eps}{(1-x)^2_{++}}
+(1-\eps)\delta(1-x)+\eps\ln\eps\,\delta'(1-x)
-\half\eps^2\,\delta''(1-x)\cr
  \frac{\eps^2}{(1-x)^3} &\to \frac{\eps^2}{(1-x)^3_{+++}}
+\half(1-\eps^2)\,\delta(1-x)-\eps(1-\eps)\,\delta'(1-x)
-\half\eps^2\ln\eps\,\delta''(1-x)\,.
}\eeq

Applying these rules to Eqs.~\re{FTr} etc., 
we obtain expressions of the form \re{Fsmalleps}.
The coefficient of $-\ln \eps $ in $\sub\cF{T,A}^{(r)}$ is the quark
splitting function $P_{qq}(x)=(1+x^2)/(1-x)$, which is singular
for $x\to 1$. The singularity is regularized by including the 
virtual contribution. As explained in Ref.~\cite{BPY}, this
term produces the logarithmic scaling violation in the quark
fragmentation function.
The second term $\sub C{0,P}^q(x)$ is the perturbative
coefficient function (in the gluon mass regularization
scheme). The remaining terms generate power corrections
of the form \re{CNP}, which we list in Sect.~4.

\subsection{Gluon fragmentation}

The characteristic functions for gluon fragmentation
depend on the quantity $\rho = \sqrt{x^2-4\eps}$ and are
defined in the phase-space region $2\sqrt\eps\leq x\leq 1+\eps$.
Naturally, they have only real gluon emission contributions.
For transverse gluon fragmentation we find
\beq\label{FTg}\eqalign{
\sub\cF T^g(x,\eps) &= \frac{4}{x}\left[
(1-x+2\eps-\eps x+\eps^2)\left(1+\frac {2\eps}{\rho^2}
\right)+\frac{1}{2}x^2\right]
\ln\left[\frac{(x+\rho)^2}{4\eps}\right] 
\cr&
-\frac{4}{\rho}(1-x+2\eps-\eps x+\eps^2)-4\rho\;,
}\eeq
and for longitudinal gluon fragmentation
\beq\label{FLg}\eqalign{
\sub\cF L^g(x,\eps) &= \frac{4}{\rho}
(1-x+2\eps-\eps x+\eps^2)
\left(1-\frac{2\eps}{\rho x}\ln\left[\frac{(x+\rho)^2}{4\eps}
\right]\right)\;.
}\eeq
There is no gluonic contribution to the asymmetric fragmentation function.

Since the upper phase-space boundary is now at $x=1+\eps$, the rule
\re{regsmalleps} for obtaining the small-$\eps$ behaviour of
expressions that are analytic at $x=1$ becomes
\beq\label{reglu}
F(x)\to F(x)+\eps [F(1)+\half\eps F'(1)]\,\delta(1-x)
+\half\eps^2 F(1)\,\delta'(1-x)\;.
\eeq
There are no gluonic contributions that are singular at $x=1$.
Instead, the lower phase-space boundary, $x=2\sqrt\eps$, is
$\eps$-dependent, and there are terms that are singular at
or near this boundary.  However, for any finite $x$ the region
of integration in Eq.~\re{convol} does not extend to the
lower phase-space boundary, and so it and the nearby
singularities are irrelevant. Thus for any finite $x$
we can safely expand $\rho = x-2\eps/x-\cdots$ and use
Eq.~\re{reglu} to obtain the small-$\eps$ limits of
Eqs.~\re{FTg} and \re{FLg}.

When taking moments of the fragmentation functions,
on the other hand, we integrate all the way down to $x=0$.
For sufficiently high moments the contribution of the
small-$x$ region is suppressed and the above procedure
will still be reliable. For lower moments, the
$x$-integration must be performed first, and then
the small-$\eps$ limit can be taken. We shall see that
the phase-space boundary and singularities at small $x$
can play a crucial r\^ole in this case. 

\mysection{Power corrections}

\subsection{Quark fragmentation}

The coefficients in Eq.~\re{CNP} for the first two power
corrections to the transverse quark coefficient function
are found from Eqs.~\re{FTq}--\re{Fv} to be
\beq\label{CTqx}\eqalign{
 \sub C{2,T}^q(x) &= \frac{4}{(1-x)_+} -2 +\frac{4}{x}
+2\,\delta(1-x)-\delta'(1-x) \cr
 \sub C{4,T}^q(x) &=  \frac{4}{(1-x)_+} +\frac{12}{x^2}
+5\,\delta(1-x)+\frac{1}{2}\,\delta''(1-x)\;.
}\eeq
The corresponding expressions in moment space,
defined by
\beq
\tilde C(N) = \int_0^1 x^{N-1}\,C(x)\,dx
\eeq
are
\beq\label{CTqN}\eqalign{
\tilde\sub C{2,T}^q(N) &= -N +3 -\frac 2N +\frac{4}{N-1}-4S_1 \cr
\tilde\sub C{4,T}^q(N) &= \frac 12 N^2 -\frac 32 N +6 +\frac{12}{N-2}-4S_1\;,
}\eeq
with
\beq\label{Sdefs}
  S_1 = \sum_{j=1}^{N-1} \frac 1j = \psi(N)+\gamma_E = \ln N+{\cal O}(1/N)\,.
\eeq

For the longitudinal quark contribution, the corresponding results are
\beq\label{CLqx}\eqalign{
 \sub C{2,L}^q(x) &= -\frac{4}{x} -2\,\delta(1-x) \cr
 \sub C{4,L}^q(x) &= -\frac{12}{x^2} -8\,\delta(1-x)-2\delta'(1-x)\;,
}\eeq
\beq\label{CLqN}\eqalign{
\tilde\sub C{2,L}^q(N) &= -2-\frac{4}{N-1}\cr
\tilde\sub C{4,L}^q(N) &= -2N -6 -\frac{12}{N-2}\;.
}\eeq

For the asymmetric coefficient function, which receives only a quark
contribution,
\beq\eqalign{
 \sub C{2,A}^q(x) &= \frac{4}{(1-x)_+} -2 +\frac{4}{x}
+2\,\delta(1-x)-\delta'(1-x) \cr
 \sub C{4,A}^q(x) &=  \frac{4}{(1-x)_+} +\frac{4}{x}
+3\,\delta(1-x)+\frac{1}{2}\,\delta''(1-x)\;,
}\eeq
\beq\eqalign{
\tilde\sub C{2,A}^q(N) &= -N +3 -\frac 2N +\frac{4}{N-1}-4S_1 \cr
\tilde\sub C{4,A}^q(N) &= \frac 12 N^2 -\frac 32 N +4 +\frac{4}{N-1}-4S_1\;.
}\eeq
Thus the $1/Q^2$ corrections to the transverse quark and asymmetric
coefficient functions are the same, but the $1/Q^4$ corrections (and higher
power corrections) are slightly different.

Note that the expressions given above for the moment coefficients
$\tilde C_{2p,\mbox{\scriptsize P}}^q$ with P=T,L are only correct
for $N>p$. As discussed earlier, for lower moments the low-$x$
singularities of the characteristic functions have to
be taken into account, and the singularity structure in $\eps$
becomes different. An important case is that of the $N=2$ moments, which
define the contributions to the transverse and longitudinal cross
sections. This will be examined more fully in Sect.~4.3.

\subsection{Gluon fragmentation}

The coefficients in Eq.~\re{CNP} for the first two power
corrections to the transverse gluon coefficient function
are found from Eq.~\re{FTg} to be
\beq\label{FTgCs}\eqalign{
 \sub C{2,T}^g(x) &= -4 +\frac{8}{x}-\frac{8}{x^2}+\frac{8}{x^3}
+2\,\delta(1-x) \cr
 \sub C{4,T}^g(x) &=  \frac{8}{x}-\frac{16}{x^2}+\frac{32}{x^3}
-\frac{64}{x^4}+\frac{64}{x^5}
+6\,\delta(1-x)+2\,\delta'(1-x)\;,
}\eeq
\beq\label{CTgN}\eqalign{
\tilde\sub C{2,T}^g(N) &= 2 -\frac 4N +\frac{8}{N-1}-\frac{8}{N-2}
+\frac{8}{N-3}\cr
\tilde\sub C{4,T}^g(N) &= 2N+4+\frac{8}{N-1}-\frac{16}{N-2}+\frac{32}{N-3}
-\frac{64}{N-4}+\frac{64}{N-5}\;.
}\eeq

For longitudinal gluon fragmentation, the corresponding results are
\beq\label{FLgCs}\eqalign{
 \sub C{2,L}^g(x) &= \frac{8}{x^2} -\frac{8}{x^3} \cr
 \sub C{4,L}^g(x) &= \frac{16}{x^2} -\frac{32}{x^3} +\frac{64}{x^4}
-\frac{64}{x^5}\;,
}\eeq
\beq\label{CLgN}\eqalign{
\tilde\sub C{2,L}^g(N) &= \frac{8}{N-2}-\frac{8}{N-3}\cr
\tilde\sub C{4,L}^g(N) &= \frac{16}{N-2}-\frac{32}{N-3}
+\frac{64}{N-4}-\frac{64}{N-5}\;.
}\eeq

The above expressions for the gluonic moment coefficients
$\tilde C_{2p,\mbox{\scriptsize P}}^g$ are only valid for $N>2p+1$.
For gluonic moments with $N\leq 2p+1$ the phase-space boundary and
low-$x$ singularities of the characteristic functions become
important and change the singularity structure in $\eps$, as
will be illustrated in detail for $N=2$ below.

\subsection{Transverse and longitudinal cross sections}

Defining the characteristic functions for the various contributions
to the cross section as $\sub\cR P^i(\eps)$ for P=T,L and $i=q,\bar q,g$,
such that
\beq\label{WDMR}\eqalign{
\sub\sigma P &= \sub\sigma P^q+\sub\sigma P^{\bar q}
+\sub\sigma P^g \>=\> 2\sub\sigma P^q+\sub\sigma P^g\;,\cr
\sub\sigma P^i &= \sigma_0\left[\frac 1 2
(\delta_{iq}+\delta_{i\bar q})\,\sub\delta{PT}
+\frac{C_F}{2\pi} \int_0^\infty
\frac{d\mu^2}{\mu^2}\,\at(\mu^2)\,
\dot\sub\cR P^i(\eps=\mu^2/Q^2)\right]}
\eeq
where $\sigma_0$ is the Born cross section, we would
expect from Eq.~\re{CNP} that the corresponding power
corrections would take the form
\beq\label{sigLNP}
\delta\sub{\sigma}{P}^i = \frac 12\sigma_0 C_F\frac{\as}{2\pi}
\int_0^\infty\frac{d\mu^2}{\mu^2} \ln\mu^2\> \delta\at(\mu^2)
\left[\tilde\sub C{2,P}^i(2)\frac{\mu^2}{Q^2}
+ \tilde\sub C{4,P}^i(2)\frac{\mu^4}{Q^4}+\cdots\right]
\eeq
where the coefficients $\tilde\sub C{2,P}^i(2)$ and
$\tilde\sub C{4,P}^i(2)$ are the $N=2$ moment coefficients
given above. This works more or less
as expected for the quark contributions, yielding $1/Q^{2p}$
power corrections, modulo logarithms. The expressions given
for $\tilde\sub C{4,P}^q(N)$ in Eqs.~\re{CTqN} and \re{CLqN}
have a pole at $N=2$; this singularity corresponds to the
logarithmic divergence of the $x$-moments of the expressions
for $\sub C{4,P}^q(x)$ in Eqs.~\re{CTqx} and \re{CLqx}. Performing
the $x$-integrations first, and then taking the small-$\eps$ limit,
we find that the quark characteristic functions are in fact
\beq\label{cRqs}\eqalign{
\sub{\cR}{T}^{q}(\eps) &= \frac 23\ln\eps+\frac{22}{9}
+\frac 32\eps^2\ln\eps(\ln\eps-1)+\frac 13\eps^3\ln\eps+\cdots\;,\cr
\sub{\cR}{L}^{q}(\eps) &= \frac 14
+3\eps\ln\eps-\frac 32\eps^2\ln\eps(\ln\eps-3)+\cdots\;,\cr
\sub{\cR}{tot}^{q}(\eps) &= \frac 23\ln\eps+\frac{97}{36}
+3\eps\ln\eps+3\eps^2\ln\eps+\frac 13\eps^3\ln\eps+\cdots\;.}
\eeq
In Eqs.~\re{cRqs}--\re{cRs}, the dots represent terms that vanish
as $\eps\to 0$, and are either analytic or $\cO(\eps^4\ln\eps)$
at $\eps=0$. Note that $\sub\cR T^q$ has a $\ln\eps$ divergence,
since the separation of $\sub\sigma T^q$ and $\sub\sigma T^g$
is not collinear safe.

The $\cO(\eps\ln\eps)$ term in $\sub{\cR}{L}^{q}$ gives a
leading power correction of order $1/Q^2$ in $\sub\sigma L^q$.
The effect of the logarithmic divergence of the $x$-moment of
$\sub C{4,P}^q(x)$ is to `promote' the $\cO(\eps\ln\eps)$
terms to $\cO(\eps\ln^2\eps)$, resulting in a $\ln Q^2$
enhancement of the $1/Q^4$ corrections to $\sub\sigma T^q$
and $\sub\sigma L^q$.

In the case of the gluon contributions, the relevant coefficients,
given by Eqs.~\re{FTgCs} and \re{FLgCs}, are so singular as $x\to 0$
that the $1/Q^{2p}$ power corrections are promoted to $1/Q$.  In
the case of $\sub\sigma L^g$, for example, we have
\beq
\frac{\mu^2}{Q^2}\int_{2\mu/Q}^{1+\mu^2/Q^2}x\,\sub C{2,L}^g(x)dx
\sim 8\frac{\mu^2}{Q^2}\int_{2\mu/Q}^1\frac{dx}{x^2}
\sim 4\frac{\mu}{Q}\;.
\eeq
Similarly, the most singular parts of $\sub C{4,L}^g(x)$
and all the higher coefficients give contributions of
order $1/Q$. Hence these terms have to be resummed,
and the true behaviour of the characteristic functions for
$\sub{\sigma}{L,T}^{g}$ involves $\sqrt\eps$ singularities.
Again performing the $x$-integrations first, and then taking the
small-$\eps$ limit, we find
\beq\label{cRgs}\eqalign{
\sub{\cR}{T}^{g}(\eps) &= -\frac 43\ln\eps-\frac{44}{9}
+\pi^2(1+\eps)^2\sqrt{\eps}
-\eps\ln\eps(\ln\eps+4)-\eps^2\ln\eps\left(\ln\eps+\frac 83\right)
+\frac{2}{15}\eps^3\ln\eps+\cdots\cr
\sub{\cR}{L}^{g}(\eps) &= 1 -\pi^2(1+\eps)^2\sqrt\eps
+\eps\ln\eps(\ln\eps-2)+\eps^2\ln\eps\left(\ln\eps-\frac{10}3\right)
-\frac{22}{15}\eps^3\ln\eps+\cdots\cr
\sub{\cR}{tot}^{g}(\eps) &= -\frac 43\ln\eps-\frac{35}{9}
-6\eps\ln\eps-6\eps^2\ln\eps-\frac 43\eps^3\ln\eps +\cdots\;.}
\eeq
Notice that the $\sqrt{\eps}$ singularities cancel in the total gluonic
contribution. Adding the quark and gluon contributions together gives
\beq\label{cRs}\eqalign{
\sub{\cR}{T}(\eps) &= \pi^2(1+\eps)^2\sqrt{\eps}
-\eps\ln\eps(\ln\eps+4)+2\eps^2\ln\eps\left(\ln\eps-\frac{17}6\right)
+\frac 45\eps^3\ln\eps+\cdots\;,\cr
\sub{\cR}{L}(\eps) &= \frac 32 -\pi^2(1+\eps)^2\sqrt{\eps}
+\eps\ln\eps(\ln\eps+4)-2\eps^2\ln\eps\left(\ln\eps-\frac{17}6\right)
-\frac{22}{15}\eps^3\ln\eps+\cdots\;,\cr
\sub{\cR}{tot}(\eps) &= \frac 32-\frac 23\eps^3\ln\eps +\cdots\;,}
\eeq
corresponding to
\beq\label{sigs}\eqalign{
\sub{\sigma}T &\simeq \sigma_0\left(1-\frac{\pi^2}{2}\frac{A_1}{Q}\right)\cr
\sub{\sigma}L &\simeq \sigma_0\left(\frac 32 C_F\frac{\as}{2\pi}
+\frac{\pi^2}{2}\frac{A_1}{Q}\right)\cr
\sub{\sigma}{tot} &\simeq \sigma_0\left(1+\frac 32 C_F\frac{\as}{2\pi}
+2\frac{\Apr_6}{Q^6}\right)\;,}
\eeq
where $\Apr_6$ is defined by Eq.~\re{adefs} with $p=3$ and
\beq\label{a1def}
A_1 \>=\>\frac{C_F}{2\pi}\int_0^\infty\frac{d\mu^2}{\mu^2}
\;\mu\;\delta\at(\mu^2)
\>=\>\frac{C_F}{\pi}\int_0^\infty d\mu\;\delta\at(\mu^2)\;.
\eeq

\mysection{Discussion}


To illustrate the above results, we have computed  the predicted $1/Q^2$
contributions to the total, transverse and longitudinal fragmentation
functions using the parametrizations provided by the ALEPH collaboration
\cite{jp_scaALEPH} for the parton fragmentation functions $D_i(x,Q^2)$.
We express the results in terms of the coefficients $\sub D{2,P}$ where
\beq\label{Ftwist}\eqalign{
\sub F P(x,Q^2) &\simeq \sub F P^\pert(x,Q^2)
\left(1+\frac{\sub D{2,P}(x,Q^2)}{Q^2}\right)\;,\cr 
\sub D{2,P}(x,Q^2) &= \frac{\Apr_2}{\sub F P^\pert(x,Q^2)}
\sum_i \int_x^1\frac{dz}{z} \sub C{2,P}^i(z)\,D_i(x/z,Q^2)\;,
}\eeq
the coefficient functions $\sub C{2,P}^i$ being as given above.
In each case, $\sub F P^\pert(x,Q^2)$ represents the relevant
lowest-order perturbative prediction. We assumed the value
$\Apr_2 = -0.2$ GeV$^2$ for the non-perturbative parameter
defined by Eq.~\re{adefs} with $p=1$.  This value is
suggested by deep inelastic data \cite{DasWeb96}.
Calculations were performed at $Q=22$ GeV, but the
$Q$-dependence of the coefficients  $\sub D{2,P}(x,Q^2)$
is small.

\begin{figure}\begin{center}
\epsfig{figure=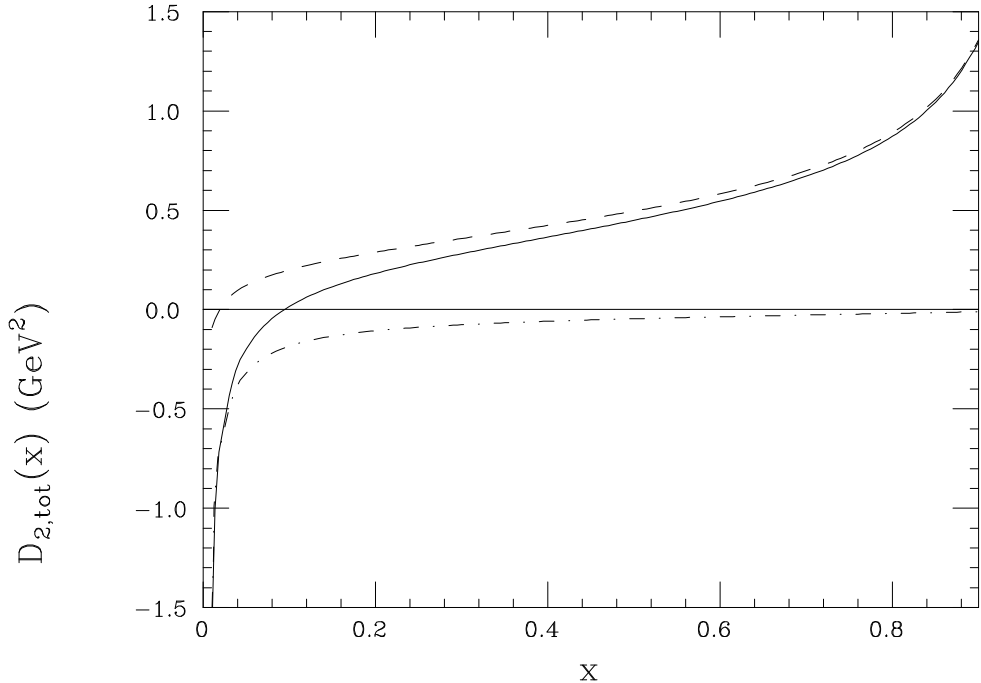,height=9.0cm}
\caption{Coefficient of $1/Q^2$ correction to the total fragmentation
function (solid), with quark (dashed) and gluon (dot-dashed) contributions
shown separately.}
\label{fig_ftotpow}
\end{center}\end{figure}
\begin{figure}\begin{center}
\epsfig{figure=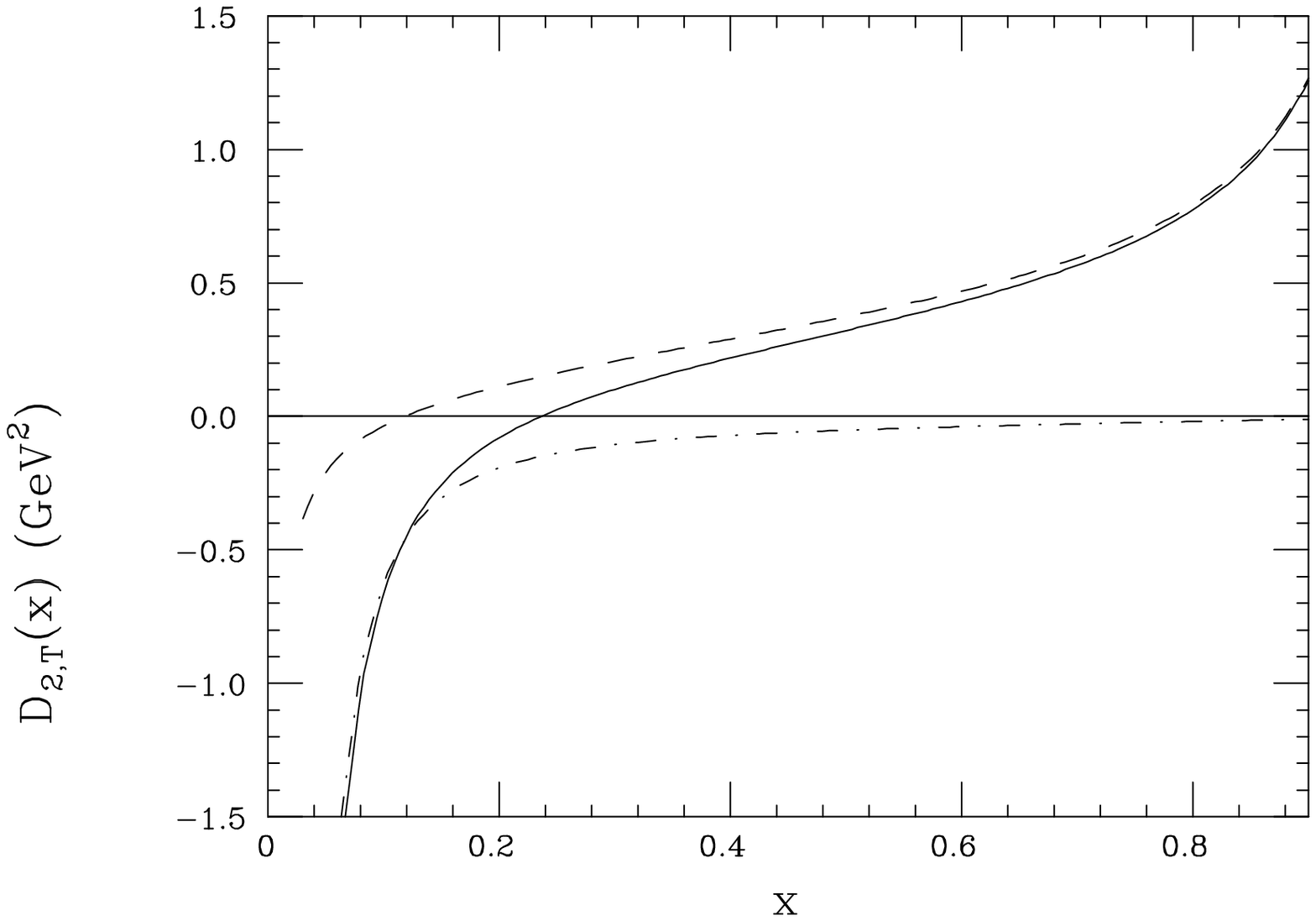,height=9.0cm}
\caption{Coefficient of $1/Q^2$ correction to the transverse fragmentation
function (solid), with quark (dashed) and gluon (dot-dashed) contributions
shown separately.}\label{fig_ftpow}
\end{center}\end{figure}
\begin{figure}\begin{center}
\epsfig{figure=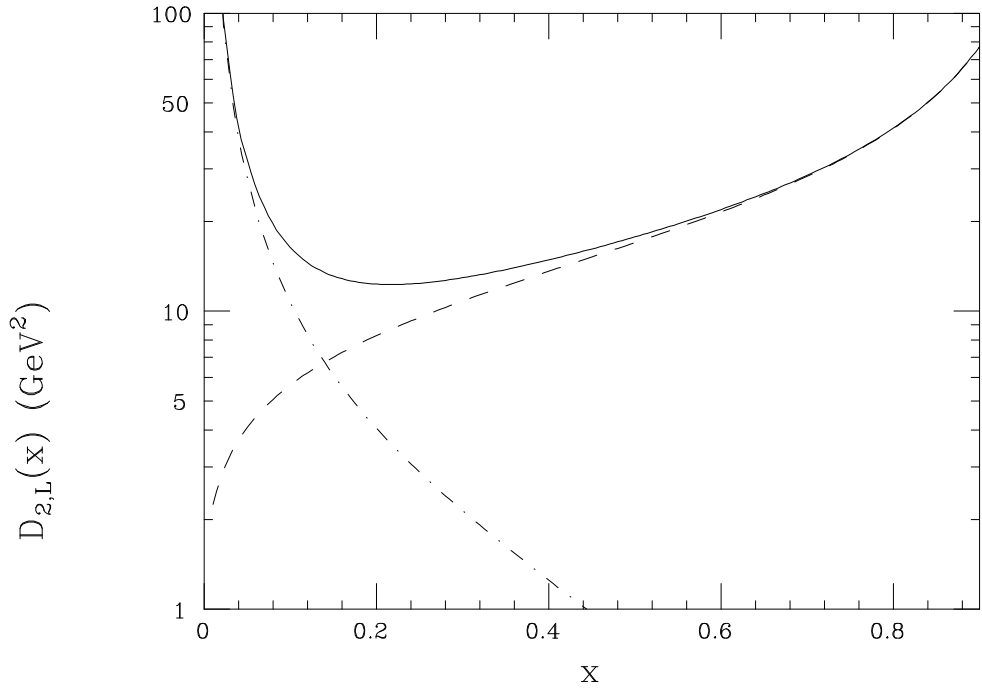,height=9.0cm}
\caption{Coefficient of $1/Q^2$ correction to the longitudinal fragmentation
function (solid), with quark (dashed) and gluon (dot-dashed) contributions
shown separately.}
\label{fig_flpow}
\end{center}\end{figure}
Figs~\ref{fig_ftotpow}, \ref{fig_ftpow} and \ref{fig_flpow}
show the results for $\sub F{tot}$, $\sub F T$ and $\sub F L$,
respectively. Note that the longitudinal coefficient $\sub D{2,L}$
is much larger than the others simply because we divide by $\sub F L^\pert$,
which is of order $\as$ relative to $\sub F{tot}^\pert$ and $\sub F T^\pert$.
We see that, as discussed above, the contributions
from gluon fragmentation give large opposing corrections to 
 $\sub F T$ and $\sub F L$ at small $x$, which tend to cancel in
$\sub F{tot}$.

Turning to the results on the transverse and longitudinal cross
sections, Eq.~\re{sigs} together with the empirical value
\re{lhad} for the $1/Q$ correction to $\sub\sigma L$ would
suggest a value $A_1\simeq 0.2$ GeV for the non-perturbative
parameter \re{a1def}. This compares fairly well with the
value $A_1\simeq 0.25$ GeV deduced event shape data in
Refs.~\citd{BPY}{Web94}. Note that in Ref.~\cite{Web94}
the coefficient of $A_1/Q$ in Eq.~\re{sigs} was given
as $2\pi\simeq 6.3$ rather than $\pi^2/2\simeq 4.9$.
This is because the calculation was performed there using
the massive-gluon phase space with a massless matrix element,
while here we include a gluon mass throughout. As discussed above,
the fact that the correction to $\sub\sigma L$ is controlled
by gluon fragmentation makes the calculation of its coefficient
by the massive gluon technique less reliable than that of
quark-dominated quantities. Nevertheless the technique is
useful in revealing the existence of the $1/Q$ correction and
its cancellation in $\sub\sigma{tot}$, and it is reassuring that
the calculations with and without mass effects in the
matrix element give numerically similar results.

\section*{Acknowledgements}

One of us (MD) is grateful to the Cavendish Laboratory and
Trinity College, Cambridge, for financial support; BRW
acknowledges the hospitality of the St Petersburg Workshop
on QCD and the CERN Theory Division while part of this work
was performed, and thanks M.\ Beneke and V.M.\ Braun for
informing him of their work on related topics \cite{BBprep}.

\par \vskip 1ex
\noindent{\large\bf References}
\begin{enumerate}
\def\cav#1{Cambridge preprint Cavendish--HEP--#1}
\def\cern#1{CERN preprint TH.#1}
\item\label{jp_scaTASSO}
       TASSO collaboration, W.\ Braunschweig \etal, \zp{41}{359}{88}.
\item\label{jp_scaDELPHI}
       DELPHI collaboration, P.\ Abreu \etal, \pl{311}{408}{93}.
\item\label{jp_scaALEPH}
       ALEPH collaboration, D.\ Buskulic \etal, \pl{357}{487}{95};\\
       \ib{364B}{247}{95} (erratum).
\item\label{jp_longOPAL}
       OPAL collaboration, R.\ Akers \etal, \zp{68}{203}{95}.
\item\label{jp_DGLAP}
       V.N.\ Gribov and L.N.\ Lipatov, \sjnp{15}{78}{72};\\
       G.\ Altarelli and G.\ Parisi, \np{126}{298}{77};\\
       Yu.L.\ Dokshitzer, \spj{46}{641}{77}.
\item\label{jp_CFP}
       G.\ Curci, W.\ Furmanski and R.\ Petronzio, \np{175}{27}{80};\\
       E.G.\ Floratos, C.\ Kounnas and R.\ Lacaze, \np{192}{417}{81}.
\item\label{jp_FP}
       W.\ Furmanski and R.\ Petronzio, \cern{2933} (1980);\\
       \pl{97}{437}{80}.
\item\label{jp_NWa}
       P.\ Nason and B.R.\ Webber, \np{421}{473}{94}.
\item\label{jp_CS}
       J.C.\ Collins and D.E.\ Soper, \ar{37}{383}{87}.
\item\label{jp_NWb}
       P.\ Nason and B.R.\ Webber, \pl{332}{405}{94}.
\item\label{GreRol}
       M.\ Greco and  S.\ Rolli, \zp{60}{169}{93};\\
       P.\ Chiappetta \etal., \np{412}{3}{94};\\
       M.\ Greco and  S.\ Rolli, \pr{52}{3853}{95};\\
       M.\ Greco, S.\ Rolli and A.\ Vicini, \zp{65}{277}{95}.
\item\label{BinKK}
       J.\ Binnewies, B.A.\ Kniehl and G.\ Kramer, \zp{65}{277}{95};\\
       \pr{52}{4947}{95}.
\item\label{jp_lp}
       EMC collaboration, M.\ Arneodo \etal, \zp{36}{527}{87};\\
       H1 collaboration, S.\ Aid \etal, \np{445}{3}{95};\\
       ZEUS collaboration, M.\ Derrick \etal, \zp{67}{93}{95}.
\item  \label{jp_had}
       CDF collaboration, F.\ Abe \etal, \prl{65}{968}{90}.
\item\label{jp_ht}
       M.\ Virchaux and A.\ Milsztajn, \pl{274}{221}{92};\\
       A.V.\ Sidorov, Dubna preprint JINR-E2-96-254 [hep-ph/9607275]. 
\item\label{jp_BalBra}
       I.I.\ Balitsky and V.M.\ Braun, \np{361}{93}{91}.
\item\label{BPY}
       Yu.L.\ Dokshitzer, G.\ Marchesini and B.R.\ Webber, \cern{95-281},\\
       to be published in {\em Nucl.\ Phys.}\ B [hep-ph/9512336].
\item\label{renormalons}
       For reviews and classic references see:\\
       V.I. Zakharov, \np{385}{452}{92};\\ 
       A.H.\ Mueller, in {\em QCD 20 Years Later}, vol.~1
       (World Scientific, Singapore, 1993).
\item\label{BBB}
       M.\ Beneke, V.M.\ Braun and V.I.\ Zakharov, \prl{73}{3058}{94};\\
       P.\ Ball, M.\ Beneke and V.M.\ Braun, \np{452}{563}{95};\\
       M.\ Beneke and V.M.\ Braun, \np{454}{253}{95}.
\item\label{BroKat}
       D.J.\ Broadhurst and A.L.\ Kataev, \pl{315}{179}{93}.
\item\label{lomax}
       C.N.\  Lovett-Turner and C.J.\ Maxwell, \np{452}{188}{95}. 
\item\label{DasWeb96}
       M.\ Dasgupta and B.R.\ Webber, \pl{382}{273}{96}.
\item\label{Stein}
       E.\ Stein, M.\ Meyer-Hermann, L.\ Mankiewicz and A.\ Sch\"afer,
       \pl{376}{177}{96};\\
       M.\ Meyer-Hermann, M.\ Maul, L.\ Mankiewicz,
       E.\ Stein and A.\ Sch\"afer, Frankfurt preprint UFTP-414-1996
       [hep-ph/9605229].
\item\label{jp_RvN}
       P.J.\ Rijken and W.L.\ van Neerven, Leiden preprint INLO-PUB-4/96
       [hep-ph/9604436].
\item\label{JETSET}
       T.\ Sj\"ostrand, \cpc{39}{347}{84};\\
       M.\ Bengtsson and T.\ Sj\"ostrand, \cpc{43}{367}{87}.
\item\label{tube}
       B.R.\ Webber, in {\em Proc.\ Summer School on Hadronic Aspects of
       Collider Physics,\\
       Zuoz, Switzerland, 1994} [hep-ph/9411384].
\item\label{Web94}
       B.R.\ Webber, \pl{339}{148}{94};\\
       Yu.L.\ Dokshitzer and B.R.\ Webber, \ib{352B}{451}{95}.
\item\label{jp_NaSe}
       P.\ Nason and M.H.\ Seymour, \np{454}{291}{95}.
\item\label{BBprep}
       M.\ Beneke, V.M.\ Braun and L.\ Magnea, paper in preparation;\\
       V.M.\ Braun, talk at ``QCD 96", Montpellier, July 1996;\\
       M.\ Beneke, talk at Workshop on Renormalons and Power
       Corrections in QCD,\\
       NORDITA, Copenhagen, August 1996.
\end{enumerate}
\end{document}